\documentstyle[amssymb,aps]{revtex}
\tightenlines 

\newtheorem{theorem}{Theorem}
\newtheorem{acknowledgement}[theorem]{Acknowledgement}

\begin{document}
\title{Achievable Luminosities at the THERA and Linac$\otimes $LHC Based ep
Colliders: 1. Round Beams}
\author{A.K. \c{C}ift\c{c}i, E. Recepo\u{g}lu}
\address{Dept. of Physics, Faculty of Sciences, Ankara University, 06100\\
Tandogan, Ankara, TURKEY}
\author{S. Sultansoy}
\address{Physics Dept., Faculty of Arts and Sciences, Gazi University, 06500,\\
Teknikokullar, Ankara, TURKEY \\
Institute of Physics, Academy of Sciences, H. Cavid Ave. 33, Baku,\\
AZERBAIJAN}
\date{\today}
\maketitle

\begin{abstract}
Different limitations on luminosity of linac-ring type colliders coming from
beam dynamics issues are considered for the examples of the $ep$ options of
the THERA\ and Linac$\otimes $LHC proposals. It is shown that $L=1.3\times
10^{31}$ for THERA\ and $L=2.7\times 10^{32}$ for Linac$\otimes $LHC can be
achieved for round beams case. Corresponding sets of parameters are given
with the explanations.
\end{abstract}

\section{INTRODUCTION}

\noindent During the last decades, linac-ring type colliders are widely
discussed concerning two purposes:

1. High luminosity particle factories \ [1],

2. High energy lepton-hadron colliders [2].

In these colliders, there are a number of unsolved problems, because of
unconventional structure with respect to well studied ring-ring and linear
colliders. There are a number of advantages which make linac-ring choice for 
$ep$ collider preferable. First of all, this is the most effective way to
reach TeV scale in $ep$ collisions. Then, it is possible to construct TeV
scale $\gamma p$ colliders on their base [3]. In addition, linacs provide
the possibility to obtain high degree of polarization of electron and
positron beams, etc.

Today, the THERA\ proposal with $\surd s\thicksim 1$ TeV [4] is the most
advanced one among the linac-ring type collider proposals and Linac$\otimes $%
LHC with $\surd s\thicksim 5$ TeV [5] should be taken into account as the
next step. In this study, we investigate limitations on parameters of the
THERA\ and Linac$\otimes $LHC $ep$ colliders coming from \ intra-beam
scattering (IBS), beam-beam tune shift, Laslett tune shift and beam beam
kink instability. At this stage, we consider round beams (both matched and
unmatched beam sizes).

\section{Fundamental Limitations on the Luminosity}

There are two most important collider parameters from physicist point of
view, namely, center of mass energy and luminosity. In addition, beam
polarization, energy spread, collision frequency and luminosity per
collision could be important in some cases. Center of mass energy is given
by $\sqrt{s}=2\sqrt{E_{e}E_{p}}$ for ultrarelativistic colliding particles.
The expression for luminosity is

\bigskip 
\begin{equation}
L=\frac{N_{e}N_{p}}{2\pi \sqrt{(\sigma _{xe}^{2}+\sigma _{xp}^{2})(\sigma
_{ye}^{2}+\sigma _{ye}^{2})}}f_{c}
\end{equation}
where $N_{e}$ is number of electrons per bunch, $N_{p}$ is number of protons
per bunch, $\sigma _{x,y}$ are horizontal and vertical beam sizes and $f_{c}$
is collision frequency. For linac-ring colliders $f_{c}=n_{b}f_{rep}$, where 
$n_{b\text{ }}$is the number of electron bunches in pulse and $f_{rep}$ is
the pulse repetition rate. To obtain high luminosity, as far as possible
high particle numbers and sufficiently small beam sizes are needed. In
addition to this, while determining this parameters, one should consider
limitations on them. Main limitations for electron beam and proton beam
parameters are explained below.

\subsection{Limitations on the electron beam parameters}

\subsubsection{Beam power}

The first restrictive limitation for electron beam is beam power

\begin{equation}
P_{e}=N_{e}E_{e}n_{b}f_{rep}
\end{equation}
which determines the maximum value of $N_{e}f_{c}$ in Eq. (1). Taking into
account the acceleration efficiency accessible value of $P_{e}$ is several
tens $MW$.

\subsubsection{Beam-beam tune shift}

The maximum number of electrons per bunch is determined by the beam-beam
tune shift limit of the proton beam

\bigskip 
\begin{equation}
\Delta Q_{p}=\frac{N_{e}r_{p}}{2\pi \gamma _{p}}\frac{\beta _{p}^{\ast }}{%
\sigma _{xe}(\sigma _{xe}+\sigma _{ye})}
\end{equation}
where $r_{p}=1.534\times 10^{-18}$ m is the classical radius of proton , $%
\gamma _{p}$ is the Lorentz factor of proton beam and $\beta _{p}^{\ast }$
is beta function at collision point. Beam-beam tune shift value generally
accepted for protons in case of ring-ring colliders is $\Delta Q\leq 0.003.$
This limit value can be a little bit larger for linac-ring type colliders.

\subsection{Limitation on the proton beam parameters}

\subsubsection{Intra-beam scattering (IBS)}

One of the most important limitations comes from intrabeam scattering. IBS
growth rates in energy spread, in the horizontal $\varepsilon _{x}$ and
vertical $\varepsilon _{y}$ emittances are defined as following [6]:

\[
\frac{1}{\tau _{p}}=\left\langle A\frac{\sigma _{h}^{2}}{\sigma _{p}^{2}}%
f(a,b,c)\right\rangle 
\]

\begin{equation}
\frac{1}{\tau _{x}}=\left\langle A\left[ f(\frac{1}{a},\frac{b}{a},\frac{c}{a%
})+\frac{D_{x}^{2}\sigma _{p}^{2}}{\sigma _{x}^{2}}f(a,b,c)\right]
\right\rangle
\end{equation}

\[
\frac{1}{\tau _{y}}=\left\langle A\left[ f(\frac{1}{b},\frac{a}{b},\frac{c}{b%
})+\frac{D_{y}^{2}\sigma _{p}^{2}}{\sigma _{y}^{2}}f(a,b,c)\right]
\right\rangle 
\]
where the brackets $\left\langle {}\right\rangle $ mean that the enclosed
quantities, combinations of beam parameters and lattice properties, are
averaged around the entire ring. In order to obtain beam size growth times
one should multiply obtained values by factor two [6]. Parameters in Eq. (4)
are:

\[
A=\frac{r_{p}^{2}cN_{p}}{64\pi ^{2}\beta ^{3}\gamma ^{4}\varepsilon
_{x}\varepsilon _{y}\sigma _{s}\sigma _{p}} 
\]

\begin{equation}
\frac{1}{\sigma _{h}^{2}}=\frac{1}{\sigma _{p}^{2}}+\frac{D_{x}^{2}\sigma
_{h}^{2}}{\sigma _{x_{\beta }}^{2}}+\frac{D_{y}^{2}\sigma _{h}^{2}}{\sigma
_{y_{\beta }}^{2}}\text{ \ \ \ \ \ \ }\sigma _{x,y}^{2}=\sigma _{x_{\beta
},y_{\beta }}^{2}+D_{x,y}^{2}\sigma _{p}^{2}
\end{equation}

\[
a=\frac{\sigma _{h}}{\gamma \sigma _{x_{\beta }^{\shortmid }}},\text{ \ }b=%
\frac{\sigma _{h}}{\gamma \sigma _{y_{\beta }^{\shortmid }}},\text{ \ }%
c=\sigma _{h}\beta \sqrt{\frac{2d}{r_{p}}} 
\]
The function $f$ is given by:

\[
f(a,b,q)=8\pi \int_{0}^{1}\left\{ 2\ln \left[ \frac{q}{2}\left( \frac{1}{P}+%
\frac{1}{Q}\right) \right] -0.577...\right\} \frac{1-3x^{2}}{PQ}dx 
\]

\begin{equation}
P^{2}=a^{2}+(1-a^{2})x^{2},\text{ \ \ }Q=b^{2}+(1-b^{2})x^{2}
\end{equation}

\[
\sigma _{x_{\beta },y_{\beta }}^{2}=\varepsilon _{x,y}\beta _{x,y\text{ }},%
\text{ \ \ \ }\sigma _{x_{\beta }^{\shortmid },y_{\beta }^{\shortmid
}}^{2}=\varepsilon _{x,y}/\beta _{x,y\text{ }} 
\]
where $\sigma _{p}$ is rms energy spread, $\sigma _{s}$ is rms bunch length, 
$\varepsilon _{x}$ and $\varepsilon _{y}$ are \ horizontal and vertical
emittances, respectively. Remaining parameters are relative velocity $\beta $%
, energy factor $\gamma $ and velocity of light $c$. $D_{x}$ and $D_{y}$ are
dispersion functions. Parameter $d$ represents a cut off for the IBS force,
which is taken equal to vertical beam size.

\subsubsection{Beam-beam kink instability}

A relative offset between the heads of the proton and electron bunches
causes to a beam-beam force, which deflects electrons. Interaction between
deflected electrons and the tail of the proton bunch causes beam-beam kick,
which can drive proton beam unstable. A stability criterion under linear
approximation is [7,8],

\begin{equation}
D_{e}\Delta Q_{p}\leq 4\nu _{s},
\end{equation}
where $D_{e}$ is disruption parameter and $\nu _{s}$ is the synchrotron tune
of the proton beam,

\subsubsection{Laslett tune shift}

Luminosity of $ep$ collisions is constrained also by Laslett tune-shift,
which leads to upper limit on the ratio $N_{p}/\varepsilon _{p}^{N}$ at
injector stage. For operating proton rings (HERA, Tevatron) and LHC injector
systems $N_{p}/\varepsilon _{p}^{N}\thickapprox 10^{17}$. However, this
limit can be overcame by a several methods; namely, higher injection energy
in the booster, smaller booster ring, appropriate cooling of the proton
beam, etc. For example, increasing of DESY III energy from 50 MeV to 120 MeV
can be achieved by moderate expenses $\sim 10$ M\$ [9]. It seems quite
realistic that $N_{p}/\varepsilon _{p}^{N}\approx $ $10^{18}$ can be
achieved by using a combination of mentioned methods.

\section{THERA}

THERA parameters given in TESLA TDR [10] are shown in the first column of \
Table I. The second column shows how to improve IBS rates for the matched
electron and proton beam sizes without changing $N_{p}/\varepsilon _{p}^{N}$
value. The third column deals with unmatched beam sizes, in which upper
limit on luminosity is imposed by beam-beam tune shift and beam beam kink
instability. Synchrotron tune for HERA is 5.12 $\times 10^{-4}$ and we have
used this value to estimate upper limit imposed by beam-beam kink
instability. The fourth column gives a possible limit on luminosity
improvement by changing injector Laslett tune shift limit with various
methods mentioned above. In this case, IBS growth time becomes main limiting
factor for matched beams, whereas beam-beam kink instability limits the
luminosity for unmatched beam case, as seen from the fifth column of the
Table I. Table I shows that in the case of matched round beams $L=1\cdot
10^{31}$ cm$^{-2}$s$^{-1}$ seems realistic for THERA and luminosity $%
L=1.3\cdot 10^{31}$ cm$^{-2}$s$^{-1}$can be achieved for unmatched round
beams.

\section{Linac$\otimes $LHC}

In this case, following [5] (see also R. Brinkmann et al. in [2]) we
consider TESLA-like accelerator as a ''linac''. A possible use of CLIC as a
''linac'' requires separate consideration, because of unmatched bunch
spacing for CLIC ($\sim ns$) and LHC ($\sim 25$ ns). In the first column of
Table II we consider LHC parameters designed for $pp$ collider option,
except bunch spacing, which is taken equal to 100 $ns$ in order to match
with TESLA\ beam structure. It is seen that $L=5\cdot 10^{30}cm^{-2}s^{-1}$
can be achieved. In next two columns LHC beam brightness is taken to be $%
N_{p}/\varepsilon _{p}^{N}=10^{17}$. We see that $L=2\cdot
10^{31}cm^{-2}s^{-1}$ can be achieved for matched beam sizes and this value
is 1.5 times greater for unmatched case. Main limitation for last case comes
from beam-beam tune shift. In difference from THERA where proton beam
brightness in the main ring is limited by IBS, Linac$\otimes $LHC allows $%
N_{p}/\varepsilon _{p}^{N}$ up to $10^{18}$ due to 7 times larger value of $%
\gamma _{p}.$ In this case, $\ L=2(2.7)\cdot 10^{32}cm^{-2}s^{-1}$ can be
achieved for matched (unmatched) beams.

\section{Conclusion}

\ It is shown that in the round beam case $L=1\cdot 10^{31}cm^{-2}s^{-1}$
for THERA\ and $L=2\cdot 10^{32}cm^{-2}s^{-1}$ for Linac$\otimes $LHC ep
options can be achieved within the reasonable upgrade of proton beam
parameters. These values may be essentially larger for flat beam case which
will be analyzed in the forthcoming paper.

\begin{acknowledgement}
\bigskip\ Authors are grateful to \"{O}. Yava\c{s} for useful discussions.
\end{acknowledgement}

\bigskip \newpage 

\bigskip 
\begin{table}[tbp] \centering%
%
\caption{ Main parameters of an ep collider based on HERA and
TESLA\label{key}}

\begin{tabular}{|l|l|l|l|l|l|}
\hline
&  & \multicolumn{2}{|l}{Upgraded THERA (1)} & \multicolumn{2}{|l|}{Upgraded
THERA (2)} \\ \hline
Parameters & TESLA TDR & matched & unmatched & matched & unmatched \\ \hline
$E_{e}$ (GeV) & 250 & 250 & 250 & 250 & 250 \\ \hline
$E_{p}$ (TeV) & 1 & 1 & 1 & 1 & 1 \\ \hline
$N_{e}$ (10$^{10}$) & 2 & 2 & 2 & 2 & 2 \\ \hline
$N_{p}$ (10$^{11}$) & 1 & 5 & 5 & 5 & 9 \\ \hline
Bunch length $\sigma _{ze}$ (mm) & 0.3 & 0.3 & 0.3 & 0.3 & 0.3 \\ \hline
Bunch length $\sigma _{zp}$ (cm) & 10 & 10 & 10 & 10 & 10 \\ \hline
$\varepsilon _{e}^{N}(\mu $m rad$)$ & 92 & 460 & 77 & 197.8 & 143.8 \\ \hline
$\varepsilon _{p}^{N}(\mu $m rad$)$ & 1 & 5 & 5 & 2.15 & 4.22 \\ \hline
Beta func. at IP $\beta _{xe,ye}^{\ast }$ (m) & 0.5 & 0.5 & 0.5 & 0.5 & 0.5
\\ \hline
Beta func. at IP $\beta _{xp,yp}^{\ast }$ (cm) & 10 & 10 & 10 & 10 & 10 \\ 
\hline
$\Delta Q_{p}$ ($10^{-3})$ & 2.44 & 0.49 & 2.94 & 1.14 & 1.56 \\ \hline
IBS $\tau _{s}/\tau _{x}$ (hour) & 1.9/2.8 & 3.1/23 & 3.1/23 & 1/3.18 & 
1.35/8.5 \\ \hline
Disruption & 0.6 & 0.6 & 0.6 & 1.4 & 1.3 \\ \hline
Bunch Spacing (ns) & 211.87 & 211.87 & 211.87 & 211.87 & 211.87 \\ \hline
Repetition rate (Hz) & 5 & 5 & 5 & 5 & 5 \\ \hline
Number of bunches in e pulse & 5600 & 5600 & 5600 & 5600 & 5600 \\ \hline
$\Delta Q_{p}D_{e}/4$ \ ($10^{-4})$ & 3.75 & 0.75 & 4.5 & 4 & 5.12 \\ \hline
$N_{p}/\varepsilon _{p}^{N}$ \ \ \ ($10^{17})$ & 1 & 1 & 1 & 2.33 & 2.13 \\ 
\hline
L$_{geo}$ ($10^{31})($\ $cm^{-2}s^{-1})$ & 0.48 & 0.48 & 0.82 & 1.1 & 1.48
\\ \hline
L(inc. Hourglass) ($10^{31})($\ $cm^{-2}s^{-1})$ & 0.43 & 0.43 & 0.7 & 1 & 
1.29 \\ \hline
\end{tabular}
\newpage 
\end{table}%
%
\newpage

\begin{table}[tbp] \centering%
%
\caption{ Main parameters of an ep collider based on Linac-LHC\label{key}}

\begin{tabular}{|l|l|l|l|l|l|}
\hline
& Designed LHC & \multicolumn{2}{|l}{Upgraded LHC (1)} & 
\multicolumn{2}{|l|}{Upgraded LHC (2)} \\ \hline
Parameters & matched & matched & unmatched & matched & unmatched \\ \hline
E$_{e}$ (TeV) & 1 & 1 & 1 & 1 & 1 \\ \hline
E$_{p}$ (TeV) & 7 & 7 & 7 & 7 & 7 \\ \hline
$N_{e}$ $(10^{10})$ & 0.7 & 0.7 & 0.7 & 0.7 & 0.7 \\ \hline
$N_{p}$ $(10^{11})$ & 1.05 & 1 & 1 & 8.6 & 20 \\ \hline
Bunch length $\sigma _{ze}$ (mm) & 1 & 1 & 1 & 1 & 1 \\ \hline
Bunch length $\sigma _{zp}$ (cm) & 7.5 & 7.5 & 7.5 & 7.5 & 7.5 \\ \hline
$\varepsilon _{e}(\mu $m rad$)$ & 46.9 & 12.5 & 3.79 & 10.75 & 10.68 \\ 
\hline
$\varepsilon _{p}(\mu $m rad$)$ & 3.75 & 1 & 1 & 0.86 & 2 \\ \hline
Beta func. at IP $\beta _{xe,ye}^{\ast }$ (m) & 2 & 2 & 2 & 2 & 2 \\ \hline
Beta func. at IP $\beta _{xp,yp}^{\ast }$ (m) & 0.1 & 0.1 & 0.1 & 0.1 & 0.1
\\ \hline
$\Delta Q_{p}$ ($10^{-3})$ & 0.24 & 0.9 & 2.97 & 1.05 & 1.05 \\ \hline
IBS $\tau _{p}/\tau _{x}$ (hour) & 108/26.3 & 23/15 & 22.9/14.9 & 2/1.26 & 
2.58/3.35 \\ \hline
Disruption & 0.22 & 0.8 & 0.08 & 8 & 8 \\ \hline
Bunch Spacing (ns) & 100 & 100 & 100 & 100 & 100 \\ \hline
Repetition rate (Hz) & 10 & 10 & 10 & 10 & 10 \\ \hline
Number of bunches in e pulse & 5000 & 5000 & 5000 & 5000 & 5000 \\ \hline
$\Delta Q_{p}D_{e}/4$ \ ($10^{-4})$ & 0.135 & 1.8 & 0.6 & 21.1 & 21.2 \\ 
\hline
$N_{p}/\varepsilon _{p}^{N}$ \ \ \ ($10^{17})$ & 0.28 & 1 & 1 & 10 & 10 \\ 
\hline
L$_{geo}$ ($10^{31})($\ $cm^{-2}s^{-1})$ & 0.6 & 2.13 & 3.23 & 21.3 & 29.5
\\ \hline
L(inc. Hourglass) ($10^{31})($\ $cm^{-2}s^{-1})$ & 0.56 & 2 & 3.03 & 20 & 
27.2 \\ \hline
\end{tabular}
\end{table}%
%

\ 

\bigskip

\end{document}